\documentclass[12pt]{article}
\usepackage{amssymb,amsmath,epsfig}

\begin{document}

\title{\bf Charged Perfect Fluid Cylindrical Gravitational Collapse}

\author{Muhammad Sharif \thanks{msharif.math@pu.edu.pk} and Ghulam Abbas
\thanks{abbasg91@yahoo.com}\\
Department of Mathematics, University of the Punjab,\\
Quaid-e-Azam Campus, Lahore-54590, Pakistan.}

\date{}
\maketitle
\begin{abstract}
This paper is devoted to study the charged perfect fluid
cylindrical gravitational collapse. For this purpose, we find a
new analytical solution of the field equations for non-static
cylindrically symmetric spacetime. We discuss physical properties
of the solution which predict gravitational collapse. It is
concluded that in the presence of electromagnetic field the
outgoing gravitational waves are absent. Further, it turns out
that when longitudinal length reduces to zero due to resultant
action of gravity and electromagnetic field, then the end state of
the gravitational collapse is a conical singularity. We also
explore the smooth matching of the collapsing cylindrical solution
to a static cylindrically symmetric solution. In this matching, we
take a special choice of constant radius of the boundary surface.
We conclude that the gravitational and Coulomb
forces of the system balance each other.\\
\end{abstract}

\section{Introduction}

Gravitational collapse of a massive star occurs when all the
thermonuclear reactions in the interior of a star could not favor
the pressure against gravity. Gravitational collapse is one of the
most important problems in general relativity. The singularity
theorems $^{1)}$ state that there exist spacetime singularities in
the realistic gravitational collapse. To investigate the nature of
spacetime singularity, Penrose  $^{2)}$ suggested a hypothesis known
as \textit{Cosmic Censorship Hypothesis} (CCH). It states that the
final fate of gravitational collapse of a massive astrophysical
object is always a black hole. This is equivalent to saying that the
singularities appearing in gravitational collapse are always clothed
by an event horizon.

Many attempts predicted that final fate of gravitational collapse
of the massive star might be a black hole or naked singularity
depending upon the choice of initial data. In this chain,
Virbhadra et al. $^{3)}$ introduced a new theoretical tool using
the gravitational lensing phenomena. In a recent paper $^{4)}$,
Virbhadra used the gravitational lensing phenomena to find an
improved form of the CCH. The classical paper of Oppenheimer and
Snyder $^{5)}$ is devoted to study dust collapse according to
which singularity is neither locally or globally naked. In other
words, the final fate of the dust collapse is a black hole. Many
people $^{6)}$ extended this work for physically existing form of
fluid with cosmological constant in spherically symmetric
background.

In order to generalize the geometry of the star, people worked on
gravitational collapse using cylindrical symmetry. The existence
of cylindrical gravitational waves provides a strong motivation in
this regard. Bronnikov and Kovalchuk $^{7)}$ were the poineers to
the work on gravitational collapse with non-spherical symmetry.
Later on, the same authors $^{8)}$ extended it for some
non-spherical exact model. Nolan $^{9)}$ investigated the naked
singularities in the cylinderical gravitational collapse of
counter rotating dust shell.

Hayward $^{10)}$ studied gravitational waves, black holes and
cosmic strings in cylindrical symmetry. Sharif and Ahmad $^{11)}$
analyzed cylindrically symmetric gravitational collapse of two
perfect fluids using the high speed approximation scheme. They
investigated the emission of gravitational radiation from
cylindrically symmetric gravitational collapse. Di Prisco et al.
$^{12)}$ discussed the shear free cylindrical gravitational
collapse using junction conditions. Nakao et al. $^{13)}$ studied
gravitational collapse of a hollow cylinder composed of dust.

Gravity is the weakest interaction among all the natural forces.
The behavior of electromagnetic field in gravitational field has
been the subject of interest for many years. Thorne $^{14)}$
developed the concept of cylindrical energy and investigated that
a strong magnetic field along the symmetry axis may halt the
cylindrical collapse of a finite cylinder before it reached to
singularity. Oron $^{15)}$ studied the relativistic magnetized
star with the poloidal and toroidal fields. Thirukkanesh and
Maharaj $^{16)}$ found that the inclusion of electromagnetic field
in gravitational collapse would counterbalance the gravitational
attraction by the Coulomb repulsive force along with pressure
gradient. In recent papers $^{17)}$-$^{19)}$, we have investigated
the effects of electromagnetic field on the perfect fluid collapse
by using junction conditions in spherically symmetric background
with positive cosmological constant. It has been found that
electromagnetic field reduces the pressure and favors the naked
singularity formation but cannot play a dominant role. Thus black
hole was formed as a final state of the gravitational collapse.

In this paper, we study the cylindrically symmetric charged
perfect fluid gravitational collapse. The main objective of this
work is to study the final fate of charged perfect fluid
gravitational collapse in the cylindrically symmetric background.
The plan of the paper is as follows: In the next section, we
discuss the solution of the Einstein-Maxwell field equations. The
physical properties of the solution are discussed in section
\textbf{3}. Section \textbf{4} gives the derivation of the
matching conditions. We summarize the results in the last section.

Geometrized units (i.e., the gravitational constant $G$=1 and speed
of light in vacuum $c=1$) are used. All the Latin and Greek indices
run from $0$ to $3$, otherwise, it will be mentioned.

\section{Solution of the Einstein Field Equations}

This section is devoted to the solution of the Einstein field
equations coupled with the charged perfect fluid as the source of
gravitation distributed per unit length of the cylinder. The
general cylindrically symmetric spacetime is given by the
following line element $^{12)}$
\begin{equation}\label{1}
ds_-^2=A^2(dt^2-dr^2)-B^2d\theta^2-C^2dz^2,
\end{equation}
where $A,~B$ and $C$ are functions of $t$ and $r$. Here we take the
following restrictions on the coordinates in order to preserve the
cylindrical symmetry of the spacetime
\begin{equation}\label{2}
-\infty\leq t \leq\infty,\quad
r\geq0,\quad0\leq\theta\leq2\pi,\quad-\infty<z<\infty.
\end{equation}
The proper unit length of the cylinder for the line element
(\ref{1}) is defined by
\begin{equation}\label{A}
l=2{\pi}{BC}.
\end{equation}
The Einstein field equations are given by
\begin{equation}\label{3}
G_{\mu}^{\nu}=\kappa(T_{\mu}^{\nu}+{T_{\mu}^{\nu}}^{(em)}).
\end{equation}
The energy-momentum tensor for perfect fluid is
\begin{equation}\label{4}
T_{{\mu}}^{\nu}={({\rho}+p)}u_{\mu}u^{\nu}-p{\delta}_{\mu}^{\nu},
\end{equation}
where $\rho$ is the energy density, $p$ is the pressure and
$u_\mu=A\delta^0_\mu$ is the four-vector velocity in co-moving
coordinates. The energy-momentum tensor for the electromagnetic
field is given by
\begin{equation}\label{5}
{T_{\mu}^{\nu}}^{(em)}=\frac{1}{4{\pi}}
(-F^{{\nu}{\lambda}}F_{{\mu}{\lambda}}+\frac{1}{4}\delta^{\nu}_{\mu}
F_{{\pi}{\lambda}}F^{{\pi}{\lambda}}),
\end{equation}
where $F_{\mu\nu}$ is the Maxwell field tensor. Now we solve the
Maxwell's field equations
\begin{eqnarray}\label{6}
F_{\mu\nu}=\phi_{\nu,\mu}-\phi_{\mu,\nu},\quad
{F^{\mu\nu}}_{;\nu}=4{\pi}J^{\mu},
\end{eqnarray}
where $\phi_{\mu}$ is the four potential and $J^{\mu}$ is the four
current. In co-moving coordinate system, the charge  per unit length
of the cylinder is assumed to be at rest so that the magnetic field
will be zero. Thus we can choose the four potential and four current
as follows
\begin{eqnarray}\label{8}
\phi_{\mu}=({\phi}(t,r),0,0,0),\quad J^{\mu}={\sigma}u^{\mu},
\end{eqnarray}
where $\sigma$ is charge density. The only non-zero component of the
field tensor is
\begin{equation}\label{9}
F_{01}=-F_{10}=-\frac{\partial\phi}{\partial {r}}.
\end{equation}
Thus the Maxwell field equations take the following form
\begin{eqnarray}\label{10}
\frac{\partial^2\phi}{\partial
{r}^2}+\frac{\partial\phi}{\partial{r}}[\frac{B'}{B}+\frac{C'}{C}-2\frac{A'}{A}]
=4{\pi}{\sigma} A^3,\\\label{11}
\frac{\partial}{\partial{t}}(\frac{1}{A^4}\frac{\partial\phi}{\partial{r}})
+(\frac{1}{A^4}\frac{\partial\phi}{\partial{r}})
[\frac{\dot{B}}{B}+\frac{\dot{C}}{C}+2\frac{\dot{A}}{A}]=0,
\end{eqnarray}
where dot and prime indicate derivatives with respect to time $t$
and radial coordinate $r$, respectively. Integration of
Eq.(\ref{10}) implies that
\begin{equation}\label{12}
\frac{\partial\phi}{\partial {r}}=\frac{2qA^2}{BC},
\end{equation}
where $q(r)=2{\pi}\int^{r}_{0}\sigma{(ABC)d{r}}$, being the
consequence of conservation law of charge, i.e., $J^\mu_{; \mu}=0$
is known as the total amount of charge per unit length of the
cylinder. It is mentioned here that Eq.(\ref{11}) is identically
satisfied by the solution of Eq.(\ref{10}). We can write
Eq.(\ref{9}) as follows
\begin{equation}\label{13}
F_{01}=-F_{10}=-\frac{2qA^2}{BC}.
 \end{equation}
The non-zero components of ${T_{\mu}^{\nu}}^{(em)}$ turn out to be
\begin{eqnarray*}
{T_{0}^{0}}^{(em)}={T_{1}^{1}}^{(em)}=-{T_{2}^{2}}^{(em)}=-{T_{3}^{3}}^{(em)}
=\frac{1}{2{\pi}}\frac{q^2}{(BC)^2}.
\end{eqnarray*}

The electric field intensity is defined by
\begin{equation}\label{14}
E(r,t)= \frac{q}{2\pi(BC)}.
\end{equation}
We assume that the charged perfect fluid distributed per unit length
of the cylinder follows along the geodesics in the interior of the
cylindrical symmetry. This requires that velocity should be uniform
and acceleration must be zero which is only possible if $A$ is
constant and in particular, we take $A=1$ (for simplicity). Thus the
field equations (\ref{3}) take the following form
\begin{eqnarray}\label{15}
-\frac{B''}{B}-\frac{C''}{C}+\frac{\dot{C}}{C}-\frac{B'C'}{BC}
+\frac{\dot{B}\dot{C}}{BC} &=&{8\pi}(\rho+{2{\pi}}E^2),
\\\label{16}
\frac{\dot{B'}}{B}+\frac{\dot{C'}}{C}&=&0,
\\\label{17}
\frac{B'C'}{BC}-\frac{\dot{B}\dot{C}}{BC}-\frac{\ddot{B}}{B}
-\frac{\ddot{C}}{C}&=&{8\pi}(p-{2{\pi}}E^2),
\\\label{18}
-\frac{\ddot{C}}{C}+\frac{C''}{C}&=&{8\pi}(p+{2{\pi}}E^2),
\\\label{19}
-\frac{\ddot{B}}{B}+\frac{B''}{B}&=&{8\pi}(p+{2{\pi}}E^2).
\end{eqnarray}
We note that there are five equations and five unknowns
$B,~C,~p,~\rho$ and $E$, thus we can find a unique solution.

For this purpose, we adopt the method of separation of variables.
The comparison of Eqs.(\ref{18}) and (\ref{19}) give
\begin{equation} \label{20}
-\frac{\ddot{C}}{C}+\frac{C''}{C}=
-\frac{\ddot{B}}{B}+\frac{B''}{B},
\end{equation}
which yields the necessary condition for pressure to be isotropic.
We take
\begin{equation}\label{21}
B= f(r)g(t),\quad C= h(r)k(t).
\end{equation}
Using Eq.(\ref{21}) in (\ref{16}), we get
\begin{equation}\label{22}
f= {\alpha}h^L, \quad k=\delta{g}^{-L},
\end{equation}
where $L~(\neq0$, for non-trivial solution) is a separation constant
while $\alpha$ and $\delta$ are integration constants. Using
Eq.(\ref{22}) in (\ref{20}), it follows that
\begin{eqnarray}\label{23}
\frac{\ddot{g}}{g}- \frac{\ddot{k}}{k}=\frac{f''}{f}-\frac{h''}{h}.
\end{eqnarray}
Since both sides are functionally independent, we put them equal to
constant say $M(\neq0$)
\begin{eqnarray}\label{24}
\frac{\ddot{g}}{g}-\frac{\ddot{k}}{k}=M=
\frac{f''}{f}-\frac{h''}{h}.
\end{eqnarray}
Application of Eq.(\ref{22}) to Eq.(\ref{24}) leads to
\begin{eqnarray}\label{25}
\frac{\ddot{g}}{g}-\frac{{\dot{g}}^2}{g^2}=\frac{M}{L+1},\quad
\frac{h''}{h}+\frac{{h'}^2}{h^2}=\frac{M}{L-1}.
\end{eqnarray}
The solution to these equations is
\begin{eqnarray}\label{27}
g(t)= {\beta}_0 \cos^\frac{1}{1-L} (Wt+t_0),\quad h(r)={\beta}_1
\cosh^\frac{1}{1+L}(Sr+r_0),
\end{eqnarray}
where ${\beta}_0,~{\beta}_1,~t_0$ and $r_0$ are constants of
integration. Further, $W$ and $S$ are given by the following
relations
\begin{eqnarray}\label{29}
W=\sqrt{\frac{M(L-1)}{L+1}},\quad S=\sqrt{\frac{M(L+1)}{L-1}}.
\end{eqnarray}

Using Eq.(\ref{27}) in (\ref{22}), it follows that
\begin{eqnarray}\label{30}
k(t)= {\beta}_2 \cos^\frac{L}{L-1} (Wt+t_0),\quad f(r)={\beta}_3
\cosh^\frac{L}{1+L}(Sr+r_0),
\end{eqnarray}
where ${\beta}_2$ and ${\beta}_3$ are constants of integration. Thus
the metric coefficients, given by Eq.(\ref{21}), turn out to be
\begin{eqnarray}\label{31}
&&B=\Omega
{\cosh^\frac{L}{1+L}(Sr+r_0)}\cos^\frac{1}{1-L}(Wt+t_0),
\\\label{32}
&&C={\Psi}{\cosh^\frac{1}{1+L}(Sr+r_0)}\cos^\frac{L}{L-1}(Wt+t_0),
\end{eqnarray}
where $\Omega=\beta_0\beta_3,~\Psi=\beta_1\beta_2$. Consequently,
the spacetime (2.1) takes the form
\begin{eqnarray}\label{33}
ds_-^2&=&dt^2-dr^2-{\Omega}^2{\cosh^\frac{2L}{1+L}
(Sr+r_0)}\cos^\frac{2}{1-L}(Wt+t_0)d\theta^2\nonumber\\
&-&{\Psi}^2{\cosh^\frac{2}{1+L}(Sr+r_0)}\cos^\frac{2L}{L-1}(Wt+t_0)dz^2.
\end{eqnarray}
Using the following transformations
\begin{eqnarray*}
Sr'=Sr+r_0, \quad Wt'=Wt+t_0,\quad
{\theta}'={\Omega}{\theta},\quad z'={\Psi}z,
\end{eqnarray*}
this metric reduces to
\begin{eqnarray}\label{34}
ds_-^2&=&dt'^2-dr'^2-{\cosh^\frac{2L}{1+L}(Sr')}\cos^\frac{2}{1-L}(Wt')d{\theta'}^2\nonumber\\
&-&{\cosh^\frac{2}{1+L}(Sr')}\cos^\frac{2L}{L-1}(Wt'){dz'}^2.
\end{eqnarray}
By assuming ${\Omega}=1$, it is clear that the above metric
preserves cylindrical symmetry with the restriction on coordinates
given by Eq.(\ref{2}). Here we take
\begin{eqnarray*}
{\tilde{B}}={\cosh^\frac{L}{1+L}(Sr')}\cos^\frac{1}{1-L}(Wt'),\quad
{\tilde{C}}={\cosh^\frac{1}{1+L}(Sr')}\cos^\frac{L}{L-1}(Wt').
\end{eqnarray*}

\section{{Physical Properties of the Solution}}

Here, we discuss some physical and geometrical properties of the
solution. The physical parameters, i.e., pressure $p$, density
$\rho$, and the electric field intensity $E$ for the metric
(\ref{34}) are given by ${\setcounter{equation}{0}}$
\begin{eqnarray}\label{a35}
p&=&\frac{1} {16\pi}[ \frac{S^2}{(1 + L)}-\frac{4{\tan}^2(W
t')W^2L}{(1 - L)^2}+\frac{W^2}{(L - 1)(2L - 1)}],\\\label{36}
E&=&[\frac{1}{32{\pi}^2}\{\frac{2L(1 + L)^2W^2
{\sec}^2(Wt')-(1+L)^3
W^2-(L-1)^3 S^2}{(1-L^2)^2}\nonumber\\
&+&\frac{2LS^2{\sec}^2h(Sr')}{(L+1)^2}\}]^\frac{1}{2},\\\label{a37}
\rho&=&\frac{1}{8\pi}[\frac{-S^2(1+L+L^2+L{\sec}^2h(Sr'))}{(L+1)^2}
+\frac{LW^2{\tan}^2(Wt')}{(L-1)^2}].
\end{eqnarray}
We would like to mention here that Eqs.(\ref{31}), (\ref{32}) and
(\ref{a35})-(\ref{a37}) satisfy all the field equations with the
restriction on constants given by Eq.(\ref{29}). The proper unit
length of the cylinder for the new metric is given by
\begin{equation} \label{38}
l=2{\pi}{\tilde{B}}{\tilde{C}}\equiv2{\pi}{\cosh(Sr')}{\cos(Wt')}
\end{equation}
and the longitudinal length in this case is
\begin{equation} \label{39}
\tilde{l}={\tilde{B}}{\tilde{C}}\equiv{\cosh(Sr')}{\cos(Wt')}.
\end{equation}

The rate of change of longitudinal length is
\begin{equation} \label{40}
\dot{\tilde{l}}=-W{\cosh(Sr')}{\sin(Wt')},
\end{equation}
where negative sign shows that motion is directed inward.$^{20)}$
Thus such motion represents gravitational collapse of the charged
perfect fluid distributed per unit length of the cylinder.

In order to analyze the nature of singularity of the solution, we use
the curvature invariants. Many scalars can be constructed from
the Riemann tensor but symmetry assumption can be used to find only
a finite number of independent scalars. Some of these are
$$ R_1=R=g^{ab}R_{ab},\quad R_2=R_{ab}R^{ab},\quad
R_3=R_{abcd}R^{abcd},\quad R_4=R^{ab}_{cd}R_{ab}^{cd}.$$ Here, we
give the analysis for the first invariant commonly known as the
Ricci scalar. For the metric (\ref{34}), it is given by
\begin{equation}\label{41}
R=\frac{2}{\tilde{{l}}}(\ddot{{\tilde{B}}}\tilde{C}-\tilde{B}\ddot{\tilde{C}}-\tilde{B}''\tilde{C}-
\tilde{C}''\tilde{B}+\dot{\tilde{B}}\dot{\tilde{C}}-\tilde{B}'\tilde{C}'),
\end{equation}
where $\tilde{l}$ is given by Eq.(\ref{39}).
We see that Ricci scalar as well as all the other curvature invariants and
physical parameters of the solution are finite for $r'\rightarrow0$. Thus
$r'=0$ is the conical singularity of the metric (\ref{34}).

Now we analyze the values of the constants for which the solution is
physical. In this solution, $L$ and $M$ are non-zero separation
constants for the non-trivial solution while the rest are
integration constants that are removed by applying the
transformations to Eq.(\ref{33}) and by evaluating the physical
parameters from the field equations. From Eq.(\ref{29}), it is clear
that the constants $W$ and $S$ cannot be chosen arbitrarily. These
are non-zero because $M\neq0$ for non-trivial solution. Further, for
$W$ and $S$ to be real, there are following four possible solutions:
\begin{eqnarray*}
&&1.\quad L<-1,\quad M>0;\quad 2.\quad L>-1,\quad M<0;\\
&&3.\quad L>1,\quad M>0;\quad ~~4.\quad L<1,\quad M<0.
\end{eqnarray*}
Keeping in mind these restrictions on the constants, we find that
the cases \textbf{1} and \textbf{2} lead to non-physical solutions
(i.e., negative energy density for the arbitrary choice of
coordinates). In the case \textbf{3}, for $0<M\leq0.5$ and
$1<L\leq1.9$, there exists a physical solution which represents the
gravitational collapse. The graphs \textbf{1-4} in this case
indicate that all the physical quantities become homogeneous. Thus
the geodesic model with charged perfect fluid distributed per unit
length of the cylinder is free of initial inhomogeneities.

It is interesting to mention here that pressure remains function
of time only for the geodesic model that is analogous to the
spherical case.$^{21)}$ In the case \textbf{4} for $0.63\leq
L\leq0.95$ and $-1<M\leq-0.10$, all quantities except pressure
behave like the case \textbf{3}, in this case pressure is negative
indicating a dark energy solution. As long as the realistic energy
condition $\rho+3p>0$ holds, the gravity remains attractive.
$^{22)}$ However, the violation of this condition i.e.,
$\rho+3p<0$ due to negative pressure, leads to the repulsive
gravitational effects. Thus in the relativistic physics, negative
pressure acting as a repulsive gravity plays the role of
preventing the gravitational collapse. We are interested to study
the gravitational collapse which is the consequence of attractive
gravity, so the case \textbf{4} is not interesting here. Thus the
only interesting case is the case \textbf{3}.

Now we proceed to discuss the energy conditions for case
\textbf{3} which are given by the following relations $^{1)}$
\begin{enumerate}
\item Weak energy condition:
\begin{equation}\label{3.3}
\rho\geq0,~~~\rho+p\geq0,
\end{equation}
\item Dominant energy condition:
\begin{equation}\label{3.4}
\rho+p\geq0,~~~\rho-p\geq0,
\end{equation}
\item Strong energy condition:
\begin{equation}\label{3.5}
\rho+3p\geq0.
\end{equation}
\end{enumerate}
For the pressure and energy density given by Eqs.(\ref{a35}) and
(\ref{a37}) respectively, it follows that
\begin{eqnarray}\label{3.6}
\rho+p&&=\frac{1}{16\pi}[\frac{S^2}{1+L}\{1-2\frac{(1+L+L^2+L{\sec}^2h(Sr'))}{1+L}\}\nonumber\\
&&+\frac{W^2}{L-1}(\frac{1}{2L-1}-2L\frac{\tan^2Wt'}{L-1})],\\\label{37}
\rho-p&&=\frac{1}{16\pi}[\frac{-S^2}{1+L}\{1+2\frac{(1+L+L^2+L{\sec}^2h(Sr'))}{1+L}\}\nonumber\\
&&+\frac{W^2}{L-1}(6L\frac{\tan^2Wt'}{L-1}-\frac{1}{2L-1})],\\\label{37}
\rho+3p&&=\frac{1}{16\pi}[\frac{S^2}{1+L}\{3-2\frac{(1+L+L^2+L{\sec}^2h(Sr'))}{1+L}\}\nonumber\\
&&+\frac{W^2}{L-1}(\frac{1}{2L-1}}-10{L\frac{\tan^2Wt'}{L-1})].
\end{eqnarray}
Notice that all these equations satisfy Eqs.(\ref{3.3})-(\ref{3.5})
for $0<M\leq0.5,~1<L\leq1.9,~0\leq{t'}\leq1$ and $0<{r'}$.

The rate of change of longitudinal length in Figure \textbf{1}
shows that the longitudinal length is a decreasing function of
time, thus the resulting solution represents the gravitational
collapse.$^{23)}$ The collapse starts at some finite time and ends
at ${t'}=1$, where longitudinal length of the cylinder reduces to
zero. Further, energy density is an increasing function of time
shown in Figure \textbf{2}. This is the strong argument for a
model to collapse. The pressure in the interior of cylinder starts
decreasing as shown in Figure \textbf{3}. This causes to initiate
the gravitational collapse, more matter is concentrated in the
small volume, hence density goes on increasing.
\begin{figure}
\center\epsfig{file=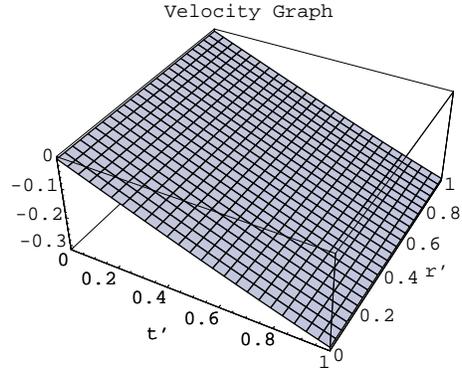, width=0.45\linewidth} \caption{Decrease
in longitudinal length with the passage of time for $0<M\leq0.5$ and
$1<L\leq1.9$ (Color online).}
\end{figure}
\begin{figure}
\center\epsfig{file=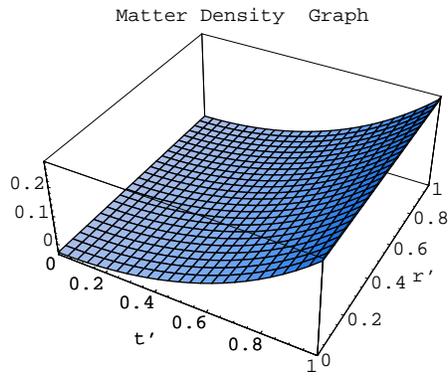, width=0.45\linewidth}\caption{Increase of
density with the passage of time for $0<M\leq0.5$ and $1<L\leq1.9$
(Color online).}
\end{figure}
\begin{figure}
\center\epsfig{file=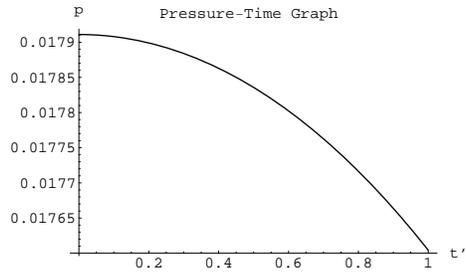, width=0.45\linewidth}\caption{Decrease in
pressure with the passage of time for $0<M\leq0.5$ and
$1<L\leq1.9$.}
\end{figure}
\begin{figure} \center\epsfig{file=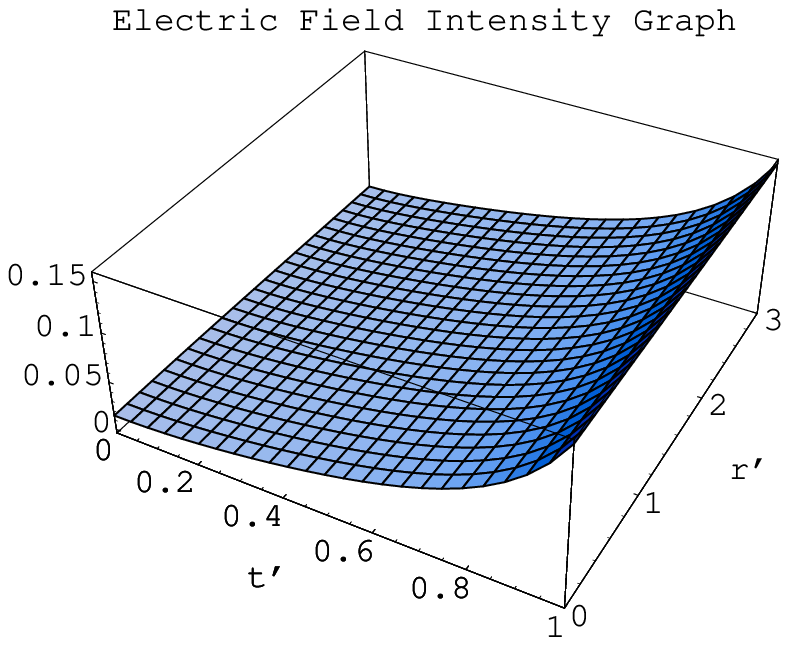, width=0.45\linewidth}
\caption{Increase of electric intensity with the passage of time for
$0<M\leq0.5$ and $1<L\leq1.9$ (Color online).}
\end{figure}

It is to be noted that decrease in the proper unit length of the
cylinder, increases the interaction between the electric charges
and a strong electromagnetic tension inside the cylinder is
created. This is an increasing function of time as shown in Figure
\textbf{4}. The coupled action of electromagnetic and
gravitational forces play a dominant role to reduce longitudinal
length of cylinder to zero.

The nature of the collapse can be seen as follows: When
latitudinal and vertical lengths of the cylinder reduce to zero,
there is a complete collapse. From the metric (2.32), we have
$g_{\theta\theta}=\tilde{B}^2,~g_{zz}=\tilde{C}^2$. Since
singularity analysis implies that the Ricci scalar diverges at a
point where the longitudinal lengths
$\tilde{l}=\tilde{B}\tilde{C}=0$. Thus when the longitudinal
length as well as the latitudinal and vertical lengths reduce to
zero, we obtain a conical singularity at ${r'}=0$.

The conical singularity has effects on the gravitational collapse of
non-zero mass objects. The conical singularity belongs to the class
of line singularities which are gravitationally weaker than point
singularities and stronger than the plane singularities. The tidal
forces for point and line singularities are so strong that these can
crush an object of finite size and non-zero mass to zero volume.
However, it was pointed out by Nakao et al. $^{24)}$ that only
conical singularity is the exceptional line singularity which does
not crush an object to zero volume that collapses onto it. Hence, it
is concluded that massive objects of finite size are collapsed on it
without crushing to zero volume. The reason of non-crushing to zero
volume does not imply that tidal forces are weak but it may be due
to the geometric structure of the conical singularity.

\section{Matching Conditions}

Following,$^{21,~25)}$ we proceed to cut the spacetime and match
it with another spacetime, which represents the exterior region of
the collapsing cylinder. We match the charged perfect fluid
solution with the electro-vacuum solution. For this purpose, we
consider the Darmois junction conditions, $^{26)}$ which require
that the first and second fundamental forms (that are the line
elements and the extrinsic curvature respectively) must be
continuous over the boundary surface ${\Sigma}$.

We assume that the $3D$ timelike boundary surface ${\Sigma}$
splits the two $4D$ cylindrically symmetric spacetimes $V^+$ and
$V^-$. The metric which describes the internal region $V^-$ is the
charged perfect fluid solution given by Eq.(\ref{34}). For the
representation of the exterior region $V^+$, a charged
cylindrically symmetric electro-vacuum solution is taken as
 $^{27)}$ ${\setcounter{equation}{0}}$
\begin{equation}\label{c2}
ds_+^2=HdT^2-\frac{1}{H}dR^2-R^2(d\theta^2+dz^2),
\end{equation}
where $H(R)=\frac{2Q^2}{R^2}-\frac{4M}{R}$, $M$ and $Q$ are the
mass and charge per unit length of the cylinder, respectively.
This choice of the exterior solution in $V^+$ region is compatible
with the charged perfect fluid solution in the interior region
$V^-$ for their smooth matching over the boundary surface
${\Sigma}$.

Now the boundary surface ${\Sigma}$ in terms of interior and
exterior coordinates can be described by the following equations
\begin{eqnarray}\label{c3}
{f}_-(r',t')=&r'-r'_{\Sigma}=0,\\\label{c4}
{f}_{+}(R,T)=&R-R_{\Sigma}(T)=0,
\end{eqnarray}
where ${r'}_{\Sigma}$ is a constant. Using these equations, the
interior and exterior metrics on $\Sigma$ take the following form
\begin{eqnarray}\label{c5}
&&(ds_-^2)_\Sigma=dt'^2-\tilde{B}^2d\theta'^2-\tilde{C}^2dz'^2,
\\\label{c6}
&&(ds_+^2)_\Sigma=[H(R_\Sigma)-\frac{1}{H(R_\Sigma)}
(\frac{dR_\Sigma}{dT})^2]dT^2-R_\Sigma^2(d\theta^2+dz^2).
\end{eqnarray}
We assume $g_{00}>0$ in Eq.(\ref{c6}) so that $T$ is a timelike
coordinate.

The continuity of the first fundamental form gives
\begin{eqnarray}\label{c7}
(\tilde{B})_\Sigma=R_\Sigma,~~~
(\tilde{C})_\Sigma=R_\Sigma,\\\label{c8}
[H(R_\Sigma)-\frac{1}{H(R_\Sigma)}
(\frac{dR_\Sigma}{dT})^2]^{\frac{1}{2}}dT=(dt')_\Sigma.
\end{eqnarray}
The components of extrinsic curvature $K^\pm_{ij}$ in terms of
interior and exterior coordinates are
\begin{eqnarray}\label{c9}
K^-_{00}&=&0,\quad
K^-_{22}=K^-_{33}=({\tilde{B}\bar{\tilde{B}}})_\Sigma,\nonumber\\
K^+_{00}&=&(R^\dag T^{\dag\dag}-T^\dag
R^{\dag\dag}-\frac{H}{2}\frac{dZ}{dR}T^{\dag\dag3}
+\frac{3}{2H}\frac{dH}{dR}T^\dag R^{\dag^2})_\Sigma,\nonumber\\
K^+_{22}&=&K^+_{33}=(HRT^\dag)_\Sigma,
\end{eqnarray}
where dagger $\dag$ and bar $\bar{}$ represent differentiation with
respect to the new coordinates $t'$ and $r'$ respectively. The
continuity of the extrinsic curvature components with Eqs.(\ref{c7})
and (\ref{c8}) leads to
\begin{eqnarray}\label{c13}
(\bar{\tilde{B}}^\dag)_\Sigma=0,\\
\label{c14}
M=[\frac{Q^2}{2\tilde{B}}+\frac{\tilde{B}}{4}
(\tilde{B}^{\dag2}-\bar{\tilde{B}}^2)]_\Sigma.
\end{eqnarray}
Here Eq.(\ref{c13}) implies that the boundary surface $\Sigma$
represents a cylinder with constant proper unit length which
behaves as boundary of the interior charged perfect fluid
distributed per unit length of the cylinder. Thus it connects the
interior charged perfect fluid solution to the exterior
electro-vacuum solution. Using this equation, Eq.(\ref{c14})
reduces to $M= ({\frac{Q^2} {2\tilde{B}}})_\Sigma$. The parametric
representation of this equation implies that the gravitational and
Coulomb forces of the system balance each other on the boundary
surface $\Sigma$. This consequence in the absence of pressure on
the boundary, is equivalent to the result found by Thirukkanesh
and Maharaj.$^{16)}$

\section{Discussion}

In this paper, we find an analytical solution to the Einstein
field equations coupled with the charged perfect fluid distributed
per unit length of the cylinder. It has been found that all the
physical parameters become homogenous but isotropic pressure is
function of time only. This property (i.e., pressure being only
time dependent) of cylindrically symmetric geodesic model is
similar to the spherical case. Further, all the energy conditions
are satisfied for a range of separation parameters and coordinates
for which the solution is physical. All sorts of singularities of
the solutions are discussed in detail. It is found that physical
singularity of the solution occurs at a point where proper areal
radius of the cylinder reduces to zero.

The physical and geometrical properties of the solution such as
increase of density and decrease in proper unit length of the
cylinder with respect to time represents the gravitational collapse.
Also, the time interval for the collapse has been investigated. It
is found that the electric field intensity of the system increases
with time. This implies that as longitudinal length of the cylinder
decreases, charges come close to each other and the interaction
between the charges is increased.

In general, it is known $^{{19),}{28)}}$ that the presence of an
electromagnetic field in a geometry causes to disturb its generic
properties which may result in the form of oscillation of
spacetime. Here, we discuss the absence of oscillation in geometry
by showing the absence of outgoing gravitational waves.

Following Pereira and Wang $^{29)}$, the component of the Weyl
tensor for our solution, is ${\Psi}_0=
-C_{\mu\nu\lambda\sigma}{L^\mu}{M^\nu}{L^\lambda}{M^\sigma}=0$,
where ${L^\mu}$ and ${M^\nu}$ are null vectors. It means that there
does not exist outgoing gravitational waves implying that
oscillations are absent in the geometry of the spacetime. Thus there
is no energy loss and hence no bouncing. Further, the absence of
fluctuations in the energy density graph \textbf{2} indicates the
absence of bounce, oscillation and gravitational waves. The
prediction that the fluctuation in energy density represents
gravitational waves is recently made by Hussain et al.$^{30)}$.

We would like to remark here that in this case electromagnetic
field is weak as compared to matter field which is obvious from
Figures \textbf{2} and \textbf{4}. This condition ($E^2<\rho$) for
the weak electromagnetic field was stated by Tasagas $^{28)}$ and
used by us $^{19)}$ to preserve the generic properties of the
Friedmann universe models. In a paper $^{29)},~{\Psi}_0\neq0$
implies that there exist outgoing gravitational waves from the
cylindrical symmetric radiating fluid gravitational collapse. It
would be interesting to find solution of the field equation with
radiating fluid and electromagnetic field in order to predict
collective role of the electromagnetic field and radiation flux
for the existence of the gravitational waves.

The Darmois criteria for the smooth matching of the cylindrically
symmetric charged perfect fluid solution to an electro-vacuum
charged static cylindrically symmetric solution leads to the
following consequences: (a) boundary surface represents a cylinder
with constant proper unit length and behaves as the boundary of the
charged perfect fluid distributed per unit length of the cylinder;
(b) the gravitational and Coulomb forces of the system balance each
other on the boundary surface $\Sigma$.

\vspace{0.25cm}

{\bf Acknowledgment}

\vspace{0.25cm}

We would like to thank the Higher Education Commission, Islamabad,
Pakistan for its financial support through the {\it Indigenous
Ph.D. 5000 Fellowship Program Batch-IV}.

\end{document}